April 18, 2024

# The Required Spatial Resolution to Assess Imbalance using Plantar Pressure Mapping

Kelsey Detels[1*], Shanni Zhou[1*], Harrison Wilson[1*], Jessica Rosendorf[1*], Ghazal Shabestanipour[1], Elias Ben Mellouk[1], David Shin[1], Joseph Schwab[1,†], Hamid Ghaednia[1,†]

**Affiliations:**

[1] Skeletal Oncology Research Group (SORG), Department of Orthopaedic Surgery, Massachusetts General Hospital, Harvard Medical School, Boston, MA 02114

[2] Center for Surgical Innovation and Engineering, Department of Orthopaedic Surgery, Cedars Sinai Healthcare System, Los Angles, CA 90048

  *-Share first authorship

  †- Share Corresponding authorship

**Corresponding Authors:**

Hamid Ghaednia, Ph.D.

hamid.ghaednia@cshs.org

Department of Orthopaedic Surgery and Department of Computational Biomedicine

Cedar Sinai Health System

Joseph Schwab, MD.

joseph.schwab@cshs.org

Department of Orthopaedic Surgery and Department of Computational Biomedicine

Cedar Sinai Health System

**Abstract**

Roughly 1/3 of adults older than 65 fall each year, resulting in more than 3 million emergency room visits, thousands of deaths, and over $50 Billion in direct costs. The Centers for Disease Control and Prevention (CDC) estimate that 1/3 of falls are preventable with effective mitigation strategies, particularly for imbalance. Therefore, quantification of imbalance is being studied extensively in recent years. In this study we investigate the feasibility of plantar pressure mapping in balance assessment through a healthy human subject study. We used an in-house plantar pressure mapping device with high precision based on Frustrated Total Internal Reflection to measure subjects sway during the Romberg test. Through the measurements obtained from all subjects, we measured the minimum spatial resolution required for plantar pressure mapping devices in assessment of balance. We conclude that most of the current devices in the market lack the requirements for imbalance measurements.

## 1. Introduction

Preserving balance in humans is a complex physiological process that requires coordination between several systems in the body, from muscle tone and bone strength to the sensory nervous systems. Disruption to any part of this process can impair balance. These disruptions can be caused by internal problems in the body such as autoimmune diseases, psychosomatic disorders, musculoskeletal abnormalities, by external factors such as physical trauma or medication, or can be simply part of our aging process.[1,2] Balance assessments are used to assist in screening, diagnosis and prognosis[3,4], as well as progress assessment following surgical and non-surgical treatment.[5] Balance assessments are not limited to one area of medicine, but are utilized most in orthopaedics[6], neurology[7], geriatric, sport medicine[8], and physical medicine and rehabilitation[4].

The majority of current balance assessment methods are subjective assessments of body sway during the Romberg test and Brunel Balance Assessment. However, the ability of these tests to assess patient balance remains limited.[9,10] Previous studies have examined the relationship between these analyses and more accurate balance capture technology, such as motion capture, plantar pressure mapping, and computerized dynamic posturography, finding that visual observation assessments fail to compete with the more sophisticated balance analysis technologies.[11] Observational assessments are subject to variations between clinicians and suffer from low replicability and reliability.[12] In contrast, balance analysis technologies attempt to quantify and identify objective measures of balance such as posture and body sway. The drawback of these technologies remains not being able to capture small changes in balance and instead capture low resolution posture, balance, and strength information on the patient. More importantly, the required accuracies needed from technologies to assess balance have not been investigated previously and remains unknown.

To mitigate these insufficiencies, clinicians additionally utilize subjective questionnaires to analyze balance. For example, the Patient-Reported Outcomes Measurement Information System (PROMIS) uses adaptive testing to longitudinally capture patient outcomes and inform the treatment process. PROMIS measures ask patients to self-report in a series of questionnaires over multiple visits.[13] Another example is the Center for Disease Control and Prevention's (CDC) Stay Independent questionnaire. This questionnaire, part of the Stopping Elderly Accidents,

Deaths & Injuries (STEADI) initiative, aims to reduce the risk of falling for older adults.[14] While patient reported outcomes and questionnaires have been shown to have some clinical validity across multiple clinical contexts, they fail to capture quantifiable information on patient balance, which are crucial for development of more sophisticated and accurate diagnosis, prognosis, and progress assessment.[15]

The clinical applications of balance measurement not only require quantification of balance but also high-accuracy or resolution of balance measurement. As an example, balance is a key indicator of healthy aging and is assessed throughout a patient's lifetime.[16] A multitude of studies have shown that adults with mild to severe Alzheimer's disease (AD) have gait and balance deficits. There is also recent evidence showing very small balances at very prior or at very early stages of dementia.[17] However, our current technologies only allow us to measure imbalances when they are high. Therefore, we are missing the transition from healthy to unhealthy balance. In this specific example of AD patients, high-resolution balance quantification could potentially enable the early detection of AD and other dementia conditions when only small signs of gait and balance deficits are present.[17,18] This raises the question: what are the minimum requirements for accurate quantification of balance? In this work we are attempting to answer this question for plantar pressure mapping based balance assessment.

Balance platforms that measure force distribution or pressure maps under feet have gained attention as a method to quantify balance more accurately than observational assessments, self-reported questionnaires, and other balance analysis technologies. In these methods, usually the center of gravity (COG), which in most studies is also called center of pressure (COP), is measured by dividing the first moment of force by the total force. It must be considered that in most of these methods, the measurements are done 30 to 100 times per second, therefore, this creates a time series of measurements. Though several different mathematical parameters can be calculated from force distribution, in most literature COP is the main studies parameter. For example, postural control is often assessed by measuring deviations in the location of the COP from an equilibrium point within the subject's base of support, Figure 1.[19] Greater movement in COP reflects increased postural sway, a widely used indication of subject balance.[20]

There are two types of technologies that can be used for measurement of COP, force plates and plantar pressure mapping (PPM). Force plates usually use four load cells to measure forces and moments between the foot and a platform, and calculate COP. However, force plates provide no information on the loading under each foot or pressure distribution at different regions of each foot, which can indicate pathologies or progression of pathologies.[21] The PPM technologies can estimate COP by measuring the pressure distribution beneath each foot, then calculating the total COP. The PPM detects COP using various sensor technologies. The most common devices available on the market utilize hundreds of pressure sensors placed underneath the foot on a mat or multiple sensors on an in-shoe system to determine the pressure under the foot at each point. These electric based devices, however, are limited in resolution to the size and placements of these sensors on the plate or in-shoe system. The highest spatial resolution available on the market for similar devices is confined to about 3.9 sensors / $cm^2$, giving a spatial resolution of about 5mm × 5mm.[22]

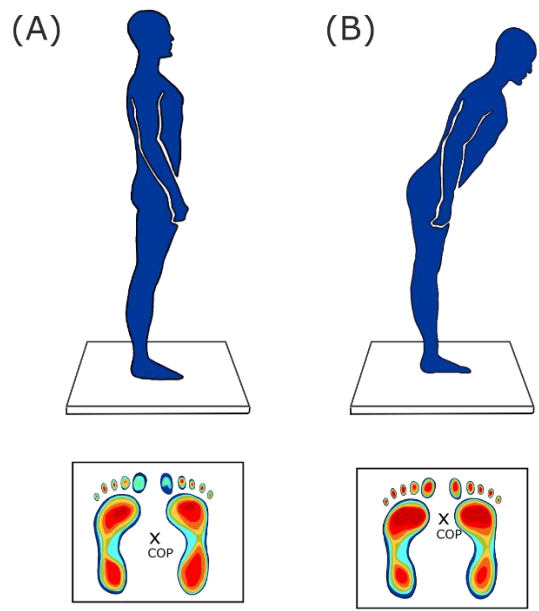

**Figure 1.** Center of pressure measurement of a subject standing on a plantar pressure mapping platform. (A) COP measurement under upright stance. (B) COP deviation under forward lean stance.

Another method of measuring plantar pressure maps that has been mostly limited to research applications is Frustrated total internal reflection (FTIR). In FTIR, light is trapped in a glass medium. At this condition, if a soft object (foot/skin) contacts the interface, the boundary condition changes and the contacting points become sources of light, enabling cameras to capture the contacting areas. Applications of FTIR in measurement of plantar pressure maps go back to 1978[23], when Betts and Duckworth used FTIR using an analogue camera to create contours of the pressure maps. This method was investigated for another decade for analyzing diabetic foot ulcerations; however, applications were limited due to the lack of physical equations required to relate light intensity values captured by the camera to the pressure values. Recently a simplified version of such equations was developed by Sharp et al.[24] that proposes the mathematical relationship between light intensity and mechanical pressure.[25] Using this equation one can measure pressure maps and therefore COP. In this work we built an in-house pressure mapping device using FTIR and Sharp et al.[24] equations to measure COP as a balance metric because it allows us to reach high spatial resolutions without losing temporal resolutions.

In this study we investigate the minimum required spatial resolution of PPM devices to quantify human balance. We do this by experimenting on healthy human subjects. We built and use an in-house FTIR based PPM device and ask 17 healthy subjects to perform a set of tests on the PPM platform including normal pose and a vestibular inhibition pose where we intentionally introduce imbalance in subjects. We record and measure PPMs of the subjects with the rate of 30 measurements/sec, then calculate the COP timeseries of each subject and for all tests. We then investigate the minimum spatial resolution required to distinguish imbalanced from balanced states. We show that to quantify balance, we need technologies that are at least 5 times better than the current sensor based available technologies. In this study we set the benchmark for future studies in quantification of imbalance using PPM. We also set the technological standard that is required for measurement of balance using PPM technologies.

## 2. Experimental Setup and Methods

For this study (Institutional Review Board protocol number 2021P001590) we recruited 17 healthy subjects. We then investigated quantification of balance based on PPM using an in-house

FTIR-based PPM device. We analyzed and compared subjects' PPMs recorded in time between normal and intentionally imbalanced conditions.

PARTICIPANTS

Seventeen neurologically healthy volunteers aged 18 to 35 years old (12 male; 5 female) participated in this study. All participants met the following criteria: 1) age 18-40 years; 2) no known neurological disorder or no known musculoskeletal injury, condition, or surgery that would affect their balance. The participants had no history of falling or concern of falling or imbalance. Participants' age, sex, height, body weight, and BMI values are shown in Table 1.

**Table 1:** Subject Demographics, Mean ± SD

|              | Total        | Male         | Female       |
| ---          | ---          | ---          | ---          |
| Number       | 17           | 12           | 5            |
| Age (years)  | 24.2 ± 5.0   | 25.7 ± 5.1   | 20.8 ± 2.3   |
| Height (cm)  | 175 ± 9      | 179 ± 6      | 165 ± 6      |
| Mass (kg)    | 74 ± 15      | 79 ± 15      | 63 ± 11      |
| BMI          | 23 ± 2       | 25 ± 4       | 24 ± 4       |

Abbreviation: BMI, body mass index (calculated as the weight in kilograms divided by the square of the height in meters).

TESTING PROTOCOL

Participants stood on the plantar pressure mapping plate for 30 seconds while completing 6 Romberg poses and one pose under vestibular inhibition:

1. Hands free and relaxed by side and eyes open
2. Hands free and relaxed by side and eyes closed
3. Hands held straight out in front and eyes open
4. Hands held straight out in front and eyes closed

5. Hands held out to the side and eyes open
6. Hands held out to the side and eyes closed
7. Hands held out to the side, head tilted back, and eyes closed (vestibular inhibition)

Participants were barefoot for all trials and before stepping on the device, we asked subjects to clean their feet with disinfectant tissues, dry with paper towel, and then to walk in place for 10 steps. The walking in place was to make sure that the mechanical properties of tissue, especially viscoelastic properties, has not been changed due to standing in place for too long. Before each patient, the platform was cleaned with disinfectant and with glass cleaner to ensure there is no dust or bacteria on the PPM device. This is especially important because dust could potentially affect the measurements. The subjects were then asked to stand still on the device for 3 minutes before the tests began. We then asked the subjects to stand as still as possible for 30 seconds for each test. For the tests that the subjects' eyes were open, the subjects were asked to stare at a marked point on the wall 8 feet in front of them. The PPMs were recorded for 30 sec by study staff for each test. The experimental setup, positions, room lighting, temperature, the experimental device and processing computers remained constant for all subjects. We repeated all the tests for each subject three times. We made sure to have at least one week between testing days for each patient. This was to reduce the effect of parameters such as sleep hours, mood, and hydration level.

DATA COLLECTION

Figure 2 shows the experimental set-up with an in-house optical pressure plate and graphical user interface application. The optical pressure plate for this study was developed using Frustrated Total Internal Reflection (FTIR). The device comprises a glass top, two RGB LED strips, a 4K USB camera, four force sensors at the corners of the glass top, a custom-made aluminum structure, an in-house image processing module for feature extraction, and an in-house easy to use Graphical User Interface for clinicians, Fig. 2. For each test, the user (study staff) instructs the participant in the proper pose, then starts the recording. Videos containing pixel intensity and color data are collected by the USB camera at 20 fps and sent to the image processing module and graphical user interface. The video is played live on the graphical user interface to confirm correct orientation and quality. The raw data were stored in the form of a video (.mp4) file and condensed file containing the force sensor values (.pkl) and then were extracted for data analysis following test completion.

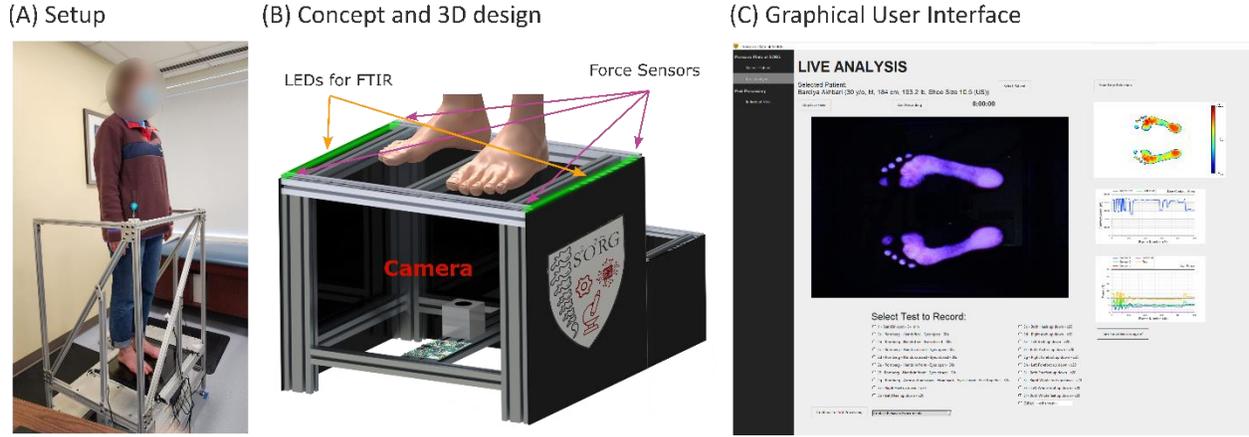

**Figure 2.** (A) Pressure mapping setup with subject standing on the device. (B) PPM system concept and 3D design. The subject stands on the glass plate, which rests on 4 force sensors and is covered by two sets of LEDs arrays on the sides for FTIR. The camera under the glass plate records the color and pixel intensity data. (C) Graphical user interface. The in-house user interface contains a live feed from the camera, and live readings from the force sensors showing force data as well as recording time, and easy recording of tests.

## DATA ANAYLSIS

To extract the data on balance from the recordings of plantar pressure maps, we conducted image segmentation and processing using the openCV library in python. After the contacting areas of the feet were segmented to remove background noise from surround background light, force/pressure measurements were extracted using light intensity-force relations established in Sharp et al.:[24]

$$I_{PP} = \kappa P^{2/3}, \qquad (1)$$

The fundamental Eq. (1) establishes a relationship between the intensity of light at each pixel, $I_{PP}$, and the pressure/force at the same pixel, $P$. The main assumptions in Eq. (1) are uniform mechanical, optical and roughness throughout both of feet. Based on Sharp's work,[24] the value $\kappa$ in Eq. (1) depends on the optical and mechanical properties of the contacting surface and can be calculated as

$$\kappa = \frac{\pi^4 A_p I_o \alpha^2 (1+\cos^2\theta_s)\phi_o}{2\lambda^3 D^2 n_o \cos\theta_r}\left[\frac{3(1-v^2)}{E}\right]^{2/3}. \qquad (2)$$

Equation 2 is based on the assumption of uniform roughness throughout subject feet,[25] where $A_p$ is the area of each pixel, $I_0$ is the incident intensity on the waveguide surface, $\alpha$ is volume polarizability of scatterers, $\phi_0$ is concentration of scatterers, $D$ is the distance between camera and contacting area, $\theta_s$ is the scattering angle, $\lambda$ is the wavelength of light trapped inside the glass top, $E$ is young modulus of the tissue, $v$ is the poisson's ratio, and

$$\cos\theta_r = \left(1 - \left(\frac{n_w}{n_0}\right)^2 \sin^2\theta_s\right)^{1/2}. \qquad (3)$$

If we consider, κ to be constant across the sole of the foot for each subject then we can use the reading of the force sensors to calculate κ. From Eq. (1) one could find P as:

$$P = \kappa I_{PP}^{3/2}, \qquad (4)$$

hence the total applied force could be calculated as

$$F_t = \int_A P dA = \kappa \int_A I_{PP}^{3/2} dA, \qquad (5)$$

where we integral the pressure over the total contact area, A to measure the total force. Therefore,

$$\kappa = \frac{F_t}{\int I_{PP}^{3/2} dA}. \qquad (6)$$

Because we measure the $F_t$ from the force sensor values designed in our system, we can easily calculate κ. To ensure that we have consistency between force and pressure measurements throughout the tests, we calculate κ at each frame based on force sensor values and then calculate the PPMs based on κ. This ensures that the total force is always consistent between the two measurements.

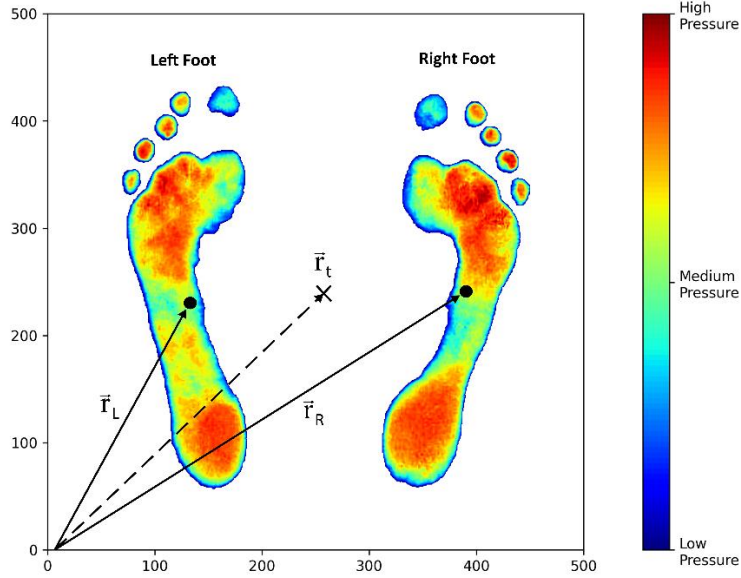

**Figure 3.** Heat map of plantar pressures under the left and right foot. Centers of pressure under the right and left feet ($\vec{r}_L$ and $\vec{r}_R$) are determined using equations (7) and (8). The total COP ($\vec{r}_t$), indicated with an x, is determined taking the pressure-weighted average of $\vec{r}_L$ and $\vec{r}_R$.

The center of pressure under each foot was evaluated over each 30 second test interval. Figure 3 demonstrates the pressure map of a single frame during one of the tests. The COP was calculated for the left, $\vec{r}_L$, and right, $\vec{r}_R$, foot by averaging the calculated pressure values and dividing by the total pressure underneath the respective foot.

$$\vec{r}_L = \frac{\sum_{A_L} P_i \vec{r}_{Li}}{\sum_{A_L} P_i} = \frac{\sum_{A_L} P_i \vec{r}_{Li}}{F_L}, \tag{7}$$

$$\vec{r}_R = \frac{\sum_{A_R} P_i \vec{r}_{Ri}}{\sum_{A_R} P_i} = \frac{\sum_{A_R} P_i \vec{r}_{Ri}}{F_R}. \tag{8}$$

The total COP, $\vec{r}_t$, is the calculated as

$$\vec{r}_t = \frac{F_L \vec{r}_L + F_R \vec{r}_R}{F_t} \tag{9}$$

where $F_L$, $F_R$, and $F_t$ are the force under left foot, force under right foot and the total force respectively. It must be considered that $\vec{r}_t$ is a 2-dimensional vector that includes both the anteroposterior (AP) and mediolateral (ML) components. We then calculate $\vec{r}_t$ for all frames of each test to create a time series of COP.

Figure 4 shows an example of COP tracking over time for one patient during one test. The top panels in figures 4a and 4b show the time series of COP plotted in 3D dimensions for tests 1 and 7 respectively. The long axis is the time dimension and horizontal and vertical axis are ML and AP components respectively. In Figure 4a, the subject is standing in the relaxed pose, tests 1, with their eyes open and their hands by their sides. In Figure 4b, the subject is standing in the vestibular inhibition pose, tests 7, with their arms out to either side, eyes closed, and head tilted back. Figure 4a & 4b, COP is plotted in 2D for the AP and ML directions, then over time for the AP and ML direction. The bottom panels show histogram plots for the distribution of COP points in the AP direction, ML direction, and in 3D.

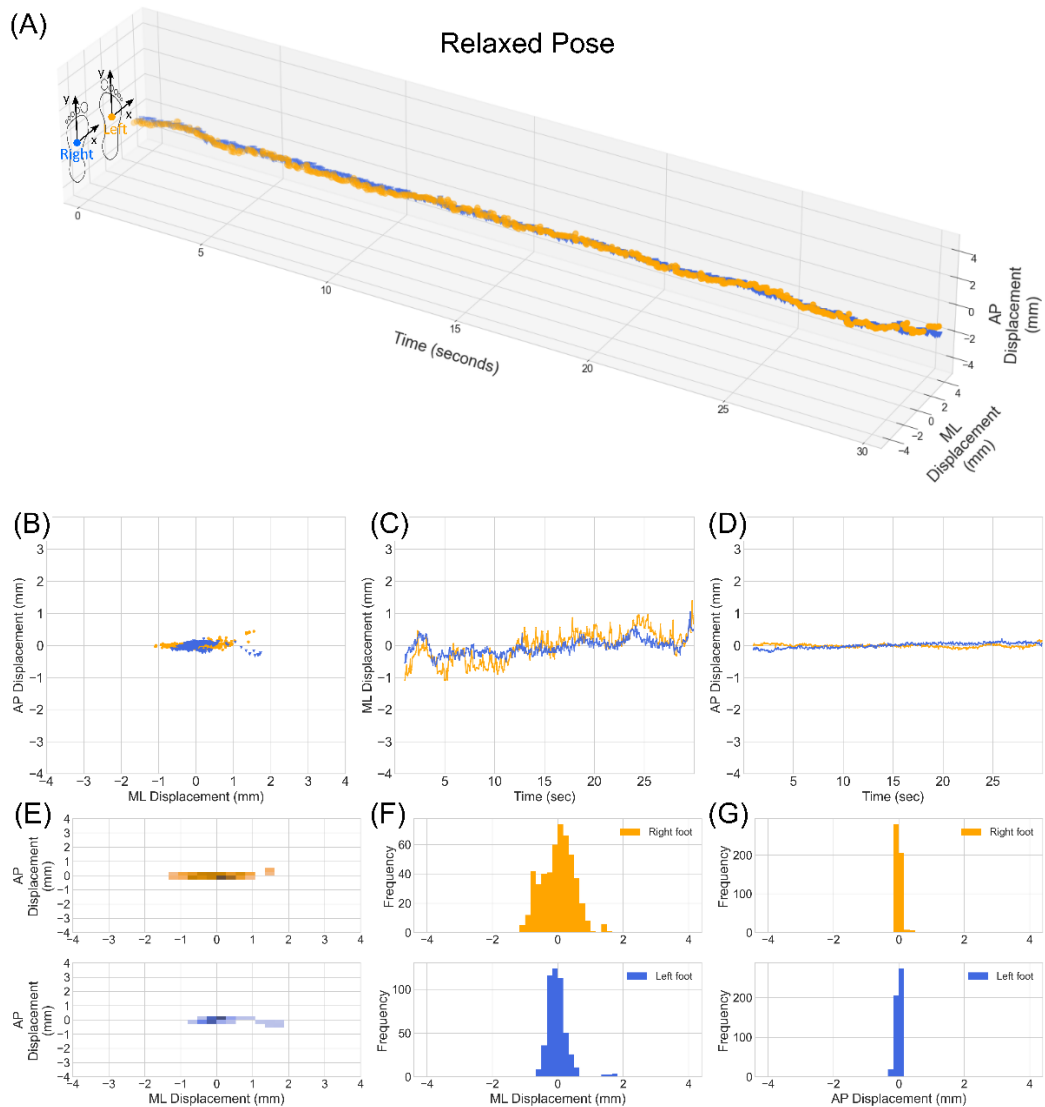

**Figure 4a.** Center of pressure tracking over time for patient in the relaxed pose. Subject standing with hands down at their sides, eyes open, feet hip width apart. (A) COP under the right (yellow) and left (blue) foot plotted in the mediolateral (ML) and anteroposterior (AP) direction over 30 seconds. (B) 2D COP trace. (B) ML displacement over time. (C) AP displacement over time. (D) Right (yellow) and left (blue) foot COP displacement 2D histogram. (E) Right (yellow) and left (blue) foot COP ML displacement histogram. (F) Right (yellow) and left (blue) foot COP AP displacement histogram.

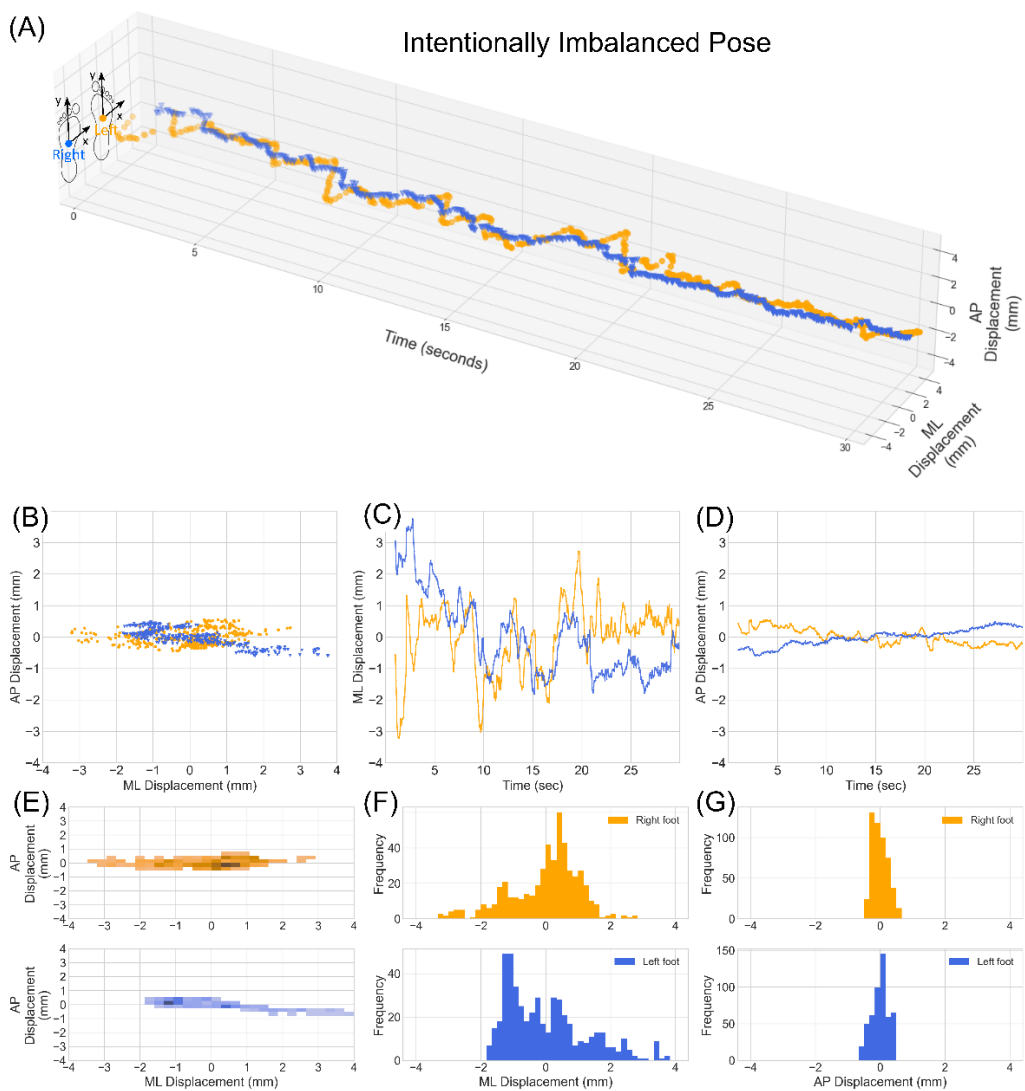

**Figure 4b.** Center of pressure tracking over time for patient while intentionally imbalanced. Subject standing with hands out to the side, eyes closed, and head tilted back. (A) COP under the right (yellow) and left (blue) foot plotted in the mediolateral (ML) and anteroposterior (AP) direction over 30 seconds. (B) 2D COP trace. (B) ML displacement over time. (C) AP displacement over time. (D) Right (yellow) and left (blue) foot COP displacement 2D histogram. (E) Right (yellow) and left (blue) foot COP ML displacement histogram. (F) Right (yellow) and left (blue) foot COP AP displacement histogram.

Various mathematical parameters can be extracted from the COP measurements taken to study imbalance. In this study, because we are solely investigating the minimum required spatial resolution for detecting imbalances we only show the standard error, SE, of the COP measurement in both AP and ML directions. This is because standard error for this problem shows the movement of the COP in time and has the dimension of length which is also similar to dimension of spatial resolution that we are studying. As an example the standard deviation of the time series cannot be used for this purpose because it is dimensionless and physically it is wrong to be used for studying the spatial resolution requirement. Therefore we calculate the standard error in the AP and ML directions in the COP measurement over time. The total standard error was then calculated as:

$$SE[\text{AP}, \text{ML}] = \frac{\sum_N |\vec{r}_t - \bar{\vec{r}}_t|}{N}, \text{ and} \tag{9}$$

$$\bar{\vec{r}}_t = \frac{\sum_N \vec{r}_t}{N}, \tag{10}$$

Where N is the number of measurements (frames) in the time series.

STATISTICAL ANALYSIS

To prove we can detect imbalance, we need to find a statistically meaningful difference between the standard error of COP for balanced and imbalanced tests. We evaluated the significances of the differences between the standard error measurements for different test groups using paired and one-sided t-tests. For the statistical analysis, our hypothesis is that the COP standard error is higher for imbalance conditions.

In order to show that for both mediolateral and anteroposterior the COP standard error is higher in imbalanced conditions, we statistically compare the standard errors between balance and imbalance tests. We performed a series of paired two-sided t-tests between test 1 and tests 2 to 6 to investigate the similarity of standard errors between them. From these paired t-tests, we reject the hypothesis that the standard error means of tests 1 through 5 are different. Thus, we grouped them together. To understand if there is any significant difference between these lumped tests and test 6 we performed a paired two-sided t-test between the grouped tests 1 through 5 and test 6. From this, we can reject the hypothesis that the mean standard error of tests 1 through 5 and test 6 are different and therefore we group tests 1 through 6. In order to conclude a significant difference between the balanced tests (grouped tests 1 to 6) and the imbalanced test (test 7), a paired one-sided t-test was then performed between grouped tests (1 through 6) and test 7. From this, we reject the hypothesis that the mean of tests 1 through 6 is similar to that of test 7. We conclude that for both mediolateral and anteroposterior directions the COP standard error for test 7 is higher than that for tests 1 through 6.

## 3. Results and Discussion

Seventeen healthy subjects stood on a PPM device for 30 seconds while completing 6 Romberg poses and one pose under vestibular inhibition. The tests were repeated three times (at least one week apart) for each patient. The tests were repeated three times for each subjects. The COP timeseries was tracked over each time period. We calculated the standard error in the AP and ML directions of the COP timeseries for each subject and each test. For each subject and test we average the SE of the COP between the three repeats. The standard error in each direction represents the variation in the movement of COP and can be used as an estimator of balance.

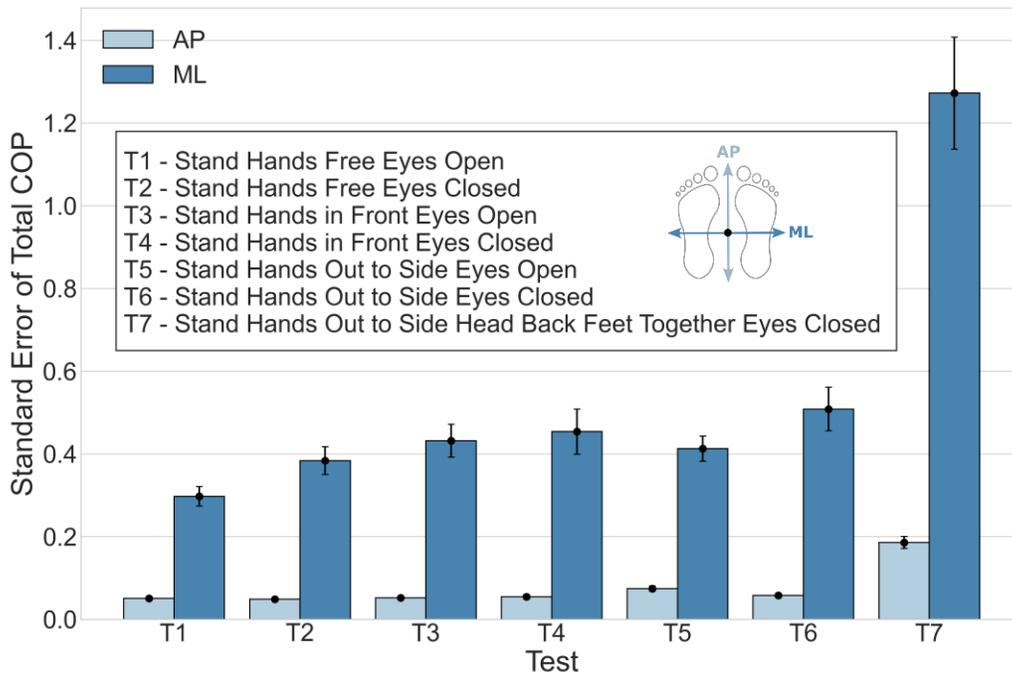

**Figure 5.** Average standard error in total COP measurements across all tests in AP and ML directions. The average standard errors (SEs) of total COP across all patients in the AP (light blue) and ML (dark blue) directions are shown. The SE of COP ranged from 0.1 - 0.2 mm and 0.3 – 1.3 mm in the AP and ML directions, respectively.

Figure 5 shows the SE of COP measurements across all tests. The SEs are averaged between all subjects for each test. The error bars show the absolute mean error/difference between the subjects. The SE of COP in the AP direction ranged from 0.1 to 0.2 mm across all tests. The SE of COP in the ML direction ranged from 0.3 to 1.3 mm across all tests. This is 16 and 4 times smaller than the spatial resolution of the current PPM technologies used in clinics for AP and ML directions respectively. The standard error in the total COP measurements for the vestibular inhibition pose, T7 (Stand Hands Out to Side Head Back Feet Together Eyes Closed), was significantly higher than all other tests in both the AP and ML directions. The standard error in the COP measurements indicate the subjects become significantly more imbalanced in test 7, as expected.

The Romberg test is often used by clinicians to examine a patient's neurological function for balance. The first pose (relaxed pose) has the subject stand with their hands by their sides, eyes open, and feet together. The subsequent poses test proprioception or the body's ability to sense its movement and location. This is done by having the patient place their hands in front of them and then to the side, both variations with their eyes open and closed. A significant increase in sway indicates a problem with the subject's proprioception and can indicate conditions such as Parkinson's disease, Wernicke's syndrome, and posterior cord syndrome. This study only includes healthy subjects who have not been diagnosed with any neurological conditions or musculoskeletal injuries that could affect their balance. Therefore, the fact that we do not observe a significant difference between COP movement in tests 1 to 6 for our subjects shows that subjects were in fact healthy with respect to their body balance. The subjects were then intentionally imbalanced in test 7 by inhibiting the vestibular system. The vestibular inhibition pose has the subject stand with their feet together, hands out to the side, eyes closed, and head tilted back. When the vestibular system is inhibited, a subjects' proprioception is limited, and they become imbalanced. In this way we are able to intentionally imbalance subjects and compare results between tests 1-6 and test 7. It must be noted that this amount of imbalance was barely noticeable by observing the subjects.

Calculating the standard error of COP time series estimates a subject's imbalance during that time. When a patient is more imbalanced, they sway more and thus their COP moves more. The standard error in COP measurements in both directions (AP and ML) ranged from 0.1 mm to 1.3 mm. The largest change was in the ML direction, with SE measurements for tests 1 through 6 around 0.3 mm and for test 7 around 1.3 mm. The change in SE between each test and test 7 was found to be significant ($p<0.05$). Thus, the 1.0 mm change in SE indicates subjects going from a balanced position to an imbalanced position. Inhibiting subjects' proprioception altered the SE in COP measurements by an average of 1.0 mm across all subjects. This indicates that PPM devices must be able to detect these small changes in COP in order to detect proprioception issues in patients.

Future work detecting imbalances in patient populations versus healthy subjects is needed to confirm clinical relevance. In this study, the vestibular inhibition pose was utilized to intentionally inhibit healthy subjects' proprioception. Data collection on patients with neurological conditions and thus inhibited proprioception systems is the next step to confirm the findings of this paper. Additionally, investigation of balance parameters on patients with multifactorial balance issues will expand the clinical use of the PPM devices to detect imbalances in all subjects for any physiological reason.

## 4. Conclusion

Intentionally unbalancing the subjects caused changes in SE of COP by less than 2 mm, indicating the need for high resolution plantar pressure mapping devices to detect changes in balance. The clinical relevance of quantifying balance via plantar pressure maps should be investigated in clinical trials in the future work. Different orthopaedic complexities should be examined separately as the effects on plantar pressure maps are largely unknown. We believe that quantifying balance with our device will advance our comprehension of related disorders and advance patient care for vulnerable populations.